\documentclass[12pt]{article}
\usepackage{natbib}
\RequirePackage{graphicx}
\RequirePackage[T1]{fontenc}
\RequirePackage{bm}
\usepackage{setspace}
\usepackage{multirow}
\usepackage{multicol}
\usepackage{url}
\usepackage{amssymb,amsmath}
\usepackage{titlesec}
\usepackage{mathtools}
\DeclarePairedDelimiter{\ceil}{\lceil}{\rceil}

\newcommand{\thetab} {{\boldsymbol{\theta}}}
\newcommand{\Sset} {\mathcal{S}}
\newcommand{\hvec} {\textbf{h}}

\newcommand{\svec} {\textbf{s}}

\newcommand{\Zvec}{\mathbf{Z}}
\newcommand{\Yvec}{\mathbf{Y}}
\newcommand{\Qvec}{\mathbf{Q}}
\newcommand{\Xvec}{\mathbf{X}}

\newcommand{\Imat} {\textbf{I}}

\newcommand{\Hmat} {\textbf{H}}

\newcommand{\intd} {\textrm{d}}
\newcommand{\epsilonb}{\boldsymbol{\varepsilon}}
\DeclareMathOperator*{\argmax}{arg\,max}

\date{}

\title{Adaptive Spatial Sampling Design for Environmental Field Prediction using Low-Cost Sensing Technologies}
\author{Eun-Hye Yoo \hspace{.2cm}\\
  Department of Geography, \\University of Buffalo, SUNY, USA \\\\
  Andrew Zammit-Mangion \hspace{.2cm}\\
  School of Mathematics and Applied Statistics, \\University of Wollongong, Australia\\\\
  Michael G. Chipeta \hspace{.2cm}\\
  Researcher in Geospatial Epidemiology, \\Big data institute, University of Oxford, UK}
\begin{document}
\bibliographystyle{agsm}
\maketitle

\begin{abstract}
The last decade has seen an explosion in data sources available for the monitoring and prediction of environmental phenomena. While several inferential methods have been developed that make predictions on the underlying process by combining these data, an optimal sampling design for when additional data is needed to complement those from other heterogeneous sources has not yet been developed. Here, we provide an adaptive spatial design strategy based on a utility function that combines both prediction uncertainty and risk-factor criteria. Prediction uncertainty is obtained through a spatial data fusion approach based on fixed rank kriging that can tackle data with differing spatial supports and signal-to-noise ratios. We focus on the application of low-cost portable sensors, which tend to be relatively noisy, for air pollution monitoring, where data from regulatory stations as well as numeric modeling systems are also available. Although we find that spatial adaptive sampling designs can help to improve predictions and reduce prediction uncertainty, low-cost portable sensors are only likely to be beneficial if they are sufficient in number and quality. Our conclusions are based on a multi-factorial simulation experiment, and on a realistic simulation of pollutants in the Erie and Niagara counties in Western New York.

\vspace{.5cm}
{\bf Keywords:} adaptive spatial sampling design,  change-of-support problem, fixed rank kriging, low-cost portable air sensors,  measurement uncertainty

\end{abstract}

\section{Introduction}

\vspace{.3cm}
The landscape of environmental science is rapidly changing due to the development of new technologies. The advances have been spurred by the decreasing cost, size, and weight, and improved reliability of environmental sensing hardware and software \citep{rundel:09}. The emergence of citizen science and robotic systems have also facilitated the application of sensing technologies to environmental science as fundamental data-gathering tools. For example, robotic systems are now used to explore deep oceans, track harmful algae blooms and pollution spread, monitor climate variables, and study remote volcanoes \citep{dunbabin:12}. Despite the increased availability of low-cost portable sensors, environmental monitoring is still hindered by incomplete observation coverage and inconsistent data quality \citep{EPA:AirSensorToolbox}.

 In the last decade, substantial efforts have been made to overcome the sparse data issues by incorporating auxiliary data from remote sensing instruments, such as measurements of Aerosol Optical Depth (AOD), or from numerical models, such as the Community Multiscale Air Quality (CMAQ) modeling system. More recently, advanced sensing technology has enabled both researchers and the community to collect real-time air quality measurements at a location of interest using low-cost portable sensors. Some of these are wearable, while others have been placed on public buses or on unmanned aerial vehicles \citep{kim:10,villa:16}. The highly portable air quality sensors are flexible and affordable, and thus allow for the deployment of scalable high-density air quality sensor networks at fine spatial and temporal scales, in both mobile and static configurations \citep{mead:13,apte:17}. These new technologies have opened a wide range of applications beyond federal and state level regulatory monitoring, although they are still in their early stages of development. Further, the accuracy of the instruments widely varies \citep{EPA:AirSensorToolbox}.

The increased availability of sensing technologies is likely to yield dramatic changes in air pollution exposure assessment, but some critical issues concerning sensor deployment, such as how many sensors are needed for a given study region and how to place sensors, still need to be resolved. There is a rich literature on environmental network design that offers some solutions; these include maximum entropy design approaches  \citep{bernardo:79,zidek:00,le:book06,fuentes:07}; spatial designs that optimally estimate the variogram model parameters \citep{bras:76,nc:90}; and designs that minimize the average kriging variance or the maximum kriging variance over a region of interest. The latter approaches can be further categorized into those that assume a known variogram model in geometric space-filling designs \citep{nychka:98,muller:book07} and those that consider spatially or temporally varying variogram models \citep{wikle:99,romary:11}. However, existing spatial and spatio-temporal sampling methods are designed for state or federal level regulatory monitoring stations (SLAMS, \url{https://www3.epa.gov/airquality/montring.html}) that yield accurate measurements and meet Federal Reference Method equivalence requirements. The quality of newly obtained air quality samples collected from low-cost portable sensors, on the other hand, may be less consistent and reliable, as shown in recent field studies \citep{castell:17,kelly:17,zimmerman:18}. Optimal designs of low-cost portable sensor networks that account for measurement uncertainty are needed for data collection.

A flexible spatial sampling approach, `adaptive sampling design,' was proposed by \cite{fanshawe:12} for spatial prediction within a model-based geostatistics framework. Recently, \cite{chipeta:16} demonstrated its application in a resource-constrained setting for survey-based malaria prevalence mapping of the community surrounding the Majete wildlife reserve, southern Malawi. An adaptive spatial or spatio-temporal sampling design selects a fixed number of sampling locations over a sequence of sampling times; at each step the data analysis is conducted with augmented sample data. The flexible adaptive design enables researchers to use a carefully selected dynamically-changing subset of monitoring locations through probabilistic models. Such models are constructed by exploiting spatial and/or temporal correlations that are present in the data and the processes under consideration. Both \cite{fanshawe:12} and \cite{chipeta:16} used model-based geostatistics, but more general spatial and spatio-temporal prediction models are also applicable.

For air quality modeling, several approaches have been proposed to integrate data obtained from disparate sources that vary in terms of their spatial resolutions and quality. Specifically, point measurements provided by regulatory air monitoring stations tend to be very accurate, while both the gridded proxy variables, such as AOD or CMAQ outputs available at a spatial resolution of 1 to 36 km, and point measurements from low-cost portable sensors, are less reliable or prone to biases and errors. One of the key challenges in air pollution data integration lies in the different spatial resolutions of the multi-source data, often referred to as the change-of-support problem. Bayesian melding \citep{fuentes:05}, for example, assumes that both observed data from the network of SLAMS and gridded proxy data arise from a common continuous latent field, which allows one to handle spatially misalignment and change-of-support \citep{gelfand:01,gotway:review,wikle:05} by taking into account the two data sources simultaneously. This approach is flexible enough to consider multiple proxy variables and multi-pollutants, although the proposed model tends to get over-parameterized very quickly \citep{chang:16}. Statistical calibration, on the other hand, uses the gridded proxy variable as a predictor in a linear model. This modeling choice considerably reduces computational burden since only a subset of proxy variables collocated with the ground observations need to be taken into account. As shown by \cite{berrocal:10} and \cite{paciorek:12}, statistical calibration yields relatively smaller prediction errors than Bayesian melding, although no missing values are allowed in the proxy variables. This is often not tenable in practice; for example, an AOD product is likely to have several missing values due to adverse weather conditions. It should be noted that, however, most solutions to the change-of-support problem have been applied only to model two data sources without low-cost portable sensor measurements whose stability and sensitivity are questionable \citep{lewis:16}.

Moreover, these statistical models that tackle the change-of-support problems in the multi-source spatial data fusion have not been used to date for an environmental network design. In the present paper, we develop a sampling design framework for monitoring long-term (months to years) concentrations of air pollutants using low-cost sensors that complement monitoring data from the network of SLAMS and gridded proxy data from CMAQ under a change-of-support model. Specifically, we use a spatial fixed rank kriging (FRK) framework \citep{cressie:08} for obtaining prediction standard error surfaces of fine scale air pollutants, which we use for optimal placement of low-cost portable sensors. The differences in the spatial resolutions of each data source are implicitly taken into account by the model, as are the uncertainties of the different measurement variables. The proposed spatial network design approach, done within a multiscale spatio-temporal data fusion framework, is general, and has the potential to be useful in a variety of application settings that aim to improve our understanding of the environment using data from low-cost portable sensors and other sources.

\section{Methods}

Consider a spatial Gaussian process $\{Y(\svec), \ \svec \in D \subset \mathbb{R}^2\}$ with mean $E(Y(\svec)) = \mu_Y(\svec)$ and covariance function $C_{\thetab}(\svec, \svec') \equiv \textrm{cov}(Y(\svec),Y(\svec') \mid {\boldsymbol \theta})$, where $\boldsymbol \theta$ is a vector of parameters and $\svec, \svec' \in D$. Here $D$ denotes the spatial domain on which the process $Y(\cdot)$ is observed through the following three types of air pollution data:

\begin{enumerate}
\item Pollutant concentration measured from regulatory instruments at $n_Z$ state and local air quality monitoring stations. We collect these measurements in $\Zvec \equiv (Y(\svec_{Z_i}) + \epsilon_{Z_i}: i = 1, \ldots, n_{Z})'$ where  $\{\epsilon_{Z_i}\}$ is a set of independent and identically distributed (i.i.d.) Gaussian measurement errors. 

\item Gridded proxy data from numerical atmospheric models, such as CMAQ, representing area-averaged concentrations of air pollution over grid cells. Let $Y_B$ denote the areal average of $Y$ over grid cell $B$, which is related to the process $Y(\cdot)$ as follows:
\begin{equation}
   Y_B \equiv Y(B) = \frac{1}{|B|} \int_{B} Y(\svec) \intd \svec. \nonumber
\end{equation}
The atmospheric model outputs are likely to be contaminated with errors associated with model parameter specification or input data. Thus, the available areal average model outputs are modeled to include measurement error. We collect the model output in $\Qvec \equiv (Y({B_i}) +  \epsilon_{Qi}: i = 1, \ldots, n_{Q})'$  where $\{\epsilon_{Qi}\}$ is a set of Gaussian measurement errors that are not necessarily uncorrelated.

\item Calibrated (i.e., unbiased) pollutant concentration measurements obtained from a low-cost portable sensor network. We collect these measurements in $\Xvec \equiv (Y(\svec_{X_i})  +  \epsilon_{X_i}: i = 1, \ldots, n_{X})'$  where $\{\epsilon_{X_i}\}$ is a set of i.i.d. Gaussian measurement errors.
\end{enumerate}
We assume that the number of portable-sensor sample sites is larger than or equal to that of the regulatory monitoring stations; that is, $n_X \geq n_Z$. We also assume that the raw data were transformed so that their empirical marginal distribution is approximately Gaussian. Note from our models that the data $\Zvec$ and $\Xvec$ are conditionally independent given the underlying process $Y(\cdot)$. This does not imply that they are marginally uncorrelated, rather that any correlation they exhibit is stemming from the  process $Y(\cdot)$.

\subsection{Adaptive Spatial Sampling Design}\label{sec:adaptsamp}

In the problem of environmental-sensor network design, the goal is to choose a finite set of locations $\Sset^X \equiv \{\svec_{X_1}, \svec_{X_2}, \ldots, \svec_{X_{n_X}}\}$ from potential sampling sites such that they are optimal, in some sense, for spatial prediction of the latent pollution field. The potential sampling locations, which are also termed ``candidate sites'' hereafter and denoted as $\Sset_0^{c} = \{\svec \in D \setminus \Sset^Z \}$, are placed within the study domain $D$, and are mutually exclusive with the fixed monitoring sites $\Sset^Z \equiv \{\svec_{Z_1}, \svec_{Z_2}, \dots, \svec_{Z_{n_Z}}\}$. The definition of candidate sites is flexible in that it can be defined either as an infinite set or a finite set of locations. The candidate sites may be chosen while taking into consideration the feasibility and the multiple objectives of the study. These objectives could include human health impact assessments \citep{baldauf:01}, near-road air quality monitoring \citep{baldauf:09}, or detection of extreme values of air pollutant concentrations \citep{chang:07a,chang:07b}. 

The algorithm we employ here closely follows that of \cite{chipeta:16}, but it is tailored to the problem of air pollutant field prediction when heterogeneous data are obtained from different sources. An initial step in the algorithm (which we reference as $k = 0$) is to obtain a surface of prediction uncertainty based on the current state of information. For this initial step we use a spatial data fusion method that is discussed in Section \ref{sec:spdf} to combine multiple spatial data sets with different uncertainty and different supports, while accounting for their spatial autocorrelation. Here, inference on the process $Y(\cdot)$ is made conditional on both ${\bf Z}$ and ${\bf Q}$, and is subsequently used to select sampling locations for low-cost portable sensors. Specifically, a new set of $b$ locations is chosen from the candidate set $\Sset^c_0$ based on the utility function that depends on both prediction uncertainty and, possibly, additional information (see Equation \eqref{eq:utility} below). This set of newly selected sampling sites in the initial step is denoted as $\Sset^{X^{(1)}}$ and the pollutant concentration measurements collected from the low-cost portable sensors at the sampling sites are denoted as ${\bf X}^{(1)}$.

If the batch size $b$ is smaller than the total number of low-cost portable sensor measurements ($b < n_X$), the adaptive sampling algorithm proceeds to step $k = 1$ by making inference on $Y(\cdot)$ using ${\bf Z}, {\bf Q}$, and ${\bf X}^{(1)}$. A new set of locations $\Sset^{X^{(2)}}$ is then chosen from the remaining locations in the candidate set $\Sset^c_1 = \Sset^c_0 \setminus \Sset^{X^{(1)}}$ based on a utility function, and the procedure is repeated until no portable sensors are left to be placed. At step $k$, the additional sampling locations are selected from the remaining candidate sites such that utility is maximized under the constraint that no two new sites in $\Sset^{X^{(k+1)}}$ are within $\delta_d$ of each other. 
This minimum distance constraint is imposed to avoid sensor network clustering within an area of high utility in the design. The adaptive sampling design is flexible in terms of the number of sampling sites selected at each step, in the sense that a different number of sampling sites $b$ can be considered at each step. Note that if the batch size is set to $1$, the sampling procedure is repeated $n_X$ times, whereas only one step is needed (the initial step) if $b = n_X$.

The algorithm proceeds as follows. Let $U_k(\svec)$ be the the utility at a location $\svec$ at step $k$ ($k=0, 1, \ldots, n_K)$, where $n_K = \ceil{\frac{n_X}{b} -1}$ with the operator $\ceil{\cdot}$ defined to round up its argument. Here, $U_0$ denotes the utility at the initial step. The utility at step $k$, $U_k$, is determined by synthesizing all the information up to step $k$, $\{{\bf Z, Q}, \Xvec_k\}$, where $\Xvec_k \equiv (\Xvec^{(1)'}, \dots, \Xvec^{(k)'})'$ and $U_0$ is  a function of only  ${\bf Z}$ and ${\bf Q}$  (since $\Xvec_0 = \Xvec^{(0)}  = \emptyset$). The new set of locations at step $k$, $\Sset^{X^{(k+1)}}$, is chosen from among all possible sets of size $b$ or less formed from locations in $\Sset^c_k$, denoted as $\tilde \Sset^c_k$, and is given by 
\begin{eqnarray}
  \Sset^{X^{(k+1)}} &=& \argmax_{\Sset \in \tilde \Sset^{c}_k} \left(\sum_{\svec \in \Sset} U_k(\svec)\right)   \label{eq:optim}\\
\textrm{subject to} && d(\svec, \svec') < \delta_d, \ \svec, \svec' \in \Sset, \nonumber 
\end{eqnarray}

\noindent where $|\Sset|$ denotes the size of $\Sset$ and $d(\svec, \svec')$ is the Euclidean distance between $\svec$ and $\svec'$. An optimal solution to Equation \eqref{eq:optim} is not generally available in closed form. We hence find a suboptimal solution via singleton adaptive sampling \citep{chipeta:16} by adding sites one by one to $\Sset^{X^{(k+1)}}$ in a greedy fashion.

In this paper we base the utility on the prediction variance in the domain of interest. However, the utility of a sampling site can also be defined as a function of other related variables that are of interest to policy makers, or it can be a mix of multiple objectives for different stakeholders. For example, \citet{kabaghe:17} use an exceedance probability criterion instead of a prediction variance given that the primary goal of the study was to delineate sub-regions where prevalence of malaria incidence is likely to exceed a policy intervention or national threshold. In the field of air pollution monitoring, the regions with elevated health risks that are potentially harmful for residents' health may require more intense monitoring than others in the study region, and thus, the utility of a location $\svec$ at step $k$ takes the form of
\begin{equation} 
U_k(\svec) = \sqrt{\textrm{var}(Y(\svec)| \Zvec, \Qvec, \Xvec_k)} + \lambda T(\svec), \label{eq:utility}
\end{equation}

\noindent where $\lambda$ is a weight that determines the relative importance of the monitoring locations $\svec$, and $T(\svec) > 0$ is a function of risk factors, such as the distance to major highways. For example, $T(\svec) = I(R(\svec) < c)$, where $R(\svec)$ is a risk function at a location $\svec$ and $c$ is a threshold.

The utility function Equation \eqref{eq:utility} is attractive as it only involves $T(\cdot)$, which is known, and the prediction variance, which for the linear Gaussian model we consider and for given $\thetab$, is only a function of the data locations and supports, and not of the data values themselves \citep[e.g.,][Chapter 2]{Rasmussen_2006}. This is an important property when designing a network for long-term air quality exposure assessment, where it would be infeasible to repeatedly wait for data to be collected prior to choosing the next set of locations in the sampling design. For this application it is appropriate to first estimate $\thetab$ based on $\Zvec$ and $\Qvec$, then obtain $\Sset^X$ through the adaptive sampling algorithm by running it from $k=0$ to $k = n_K$, and then deploying the sensors in one go.

\subsection{Spatial Data Fusion}\label{sec:spdf} 

The evaluation of the utility function in Equation \eqref{eq:utility} requires the computation of the predictive variance at arbitrary locations using data from multiple sources with different spatial supports and unequal measurement errors. The computation of this prediction variance needs to be computationally efficient since it is used iteratively in an optimization routine. A spatial modeling framework that can easily accommodate the change-of-support problem and the computational requirement is fixed rank kriging (FRK). Here, we briefly give an outline of the computational framework; more details can be found in \citet{zammit:18}.

Consider, first, the process $Y(\cdot)$. In FRK, $Y(\cdot)$ is modeled as a sum of (large-scale) fixed effects, $r$ pre-specified basis functions weighted by random coefficients, and a fine-scale variation term. Specifically,
\begin{equation}
Y(\svec) = {\bf x}(\svec)' \beta + {\boldsymbol \phi}(\svec)' {\boldsymbol \eta} + \xi(\svec), \quad \svec \in D,
\label{eq:Y}
\end{equation}

\noindent where the large-scale variation (or trend) is captured through ${\bf x}(\svec)'{\boldsymbol \beta}$ with known $p$-dimensional covariate vector ${\bf x}(\svec) \equiv (x_1(\svec), \ldots, x_p(\svec))',~\svec \in D$; the medium-scale variation is captured through the $r$ basis functions ${\boldsymbol \phi}(\svec) \equiv (\phi_1(\svec), \ldots, \phi_r(\svec))',~\svec \in D$, and the fine-scale variation is captured through $\xi(\svec),~\svec \in D,$ which is almost spatially uncorrelated. The random vector of coefficients ${\boldsymbol \eta} = (\eta_1, \ldots, \eta_r)'$ have zero mean and covariance matrix $\bf K$, which models the dependence between the basis functions. The basis functions are local with compact support and multi-resolutional: basis-function coefficients across multiple resolutions are modeled as independent, while those in the same resolution have a covariance that decreases exponentially with the distance between the basis-function centroids; see \cite{zammit:18} for further details. The process $\phi(\cdot)'\eta + \xi(\cdot)$ is sometimes referred to as a spatial random effects model.

To cater for change-of-support, it is convenient to consider the process $Y(\cdot)$ on a fine lattice composed of $N$ (where $N$ is large) small areas $\{A_i: i =1,\dots, N\}$ defined over $D$, which are often termed \emph{basic areal units} (BAUs).  The process evaluated over the BAUs, which we collect in the vector ${\bf Y} \equiv (Y(A_i): i= 1,\ldots, N)'$, is then given by

\begin{equation}
Y_i \equiv Y(A_i) = \frac{1}{|A_i|} \int_{A_i} Y(\svec) \textrm{d} \svec = {\bf x}_i' {\boldsymbol \beta} + \phi_i' {\boldsymbol \eta} + \xi_i,
\end{equation}
fir $i = 1,\dots,N$, where
$$
{\bf x}_i = \left ( \frac{1}{|A_i|} \int_{A_i}x_l(\svec) \textrm{d} \svec: l=1, \ldots, p \right)',
$$
is a $(p \times 1)$ vector of covariates formed by averaging the covariates over the BAUs,
$$\phi'_i = \left(\frac{1}{|A_i|} \int_{A_i} \phi_l(\svec) \textrm{d}\svec: l = 1,\dots, r \right)',$$
is the vector of basis functions averaged over the BAUs, and
$$\xi_i = \frac{1}{|A_i|}\int_{A_i}\xi(\svec) \textrm{d}\svec,$$
is a fine-scale variation term which is assumed to be spatially uncorrelated and normally distributed with mean $0$ and variance $\sigma^2_\xi$. 

Consider the vector of \emph{point-referenced}  data ${\bf Z} = (Z(\svec_{j}) : j = 1, \ldots, n_{Z})'$. In the BAU setting, we assume that each data point is a noisy observation of just one BAU. Hence, we can write
\begin{equation}
\textbf{Z} = \textbf{H}_Z\mathbf{Y} + \boldsymbol{\varepsilon}_Z,
\end{equation}
\noindent where $\boldsymbol{\varepsilon}_Z$ is normally distributed with mean $\mathbf{0}$ and covariance matrix $\sigma^2_{\boldsymbol{\varepsilon}_Z}\textbf{I}$, and $\textbf{H}_Z$ is an \emph{incidence} matrix that contains one non-zero per row. Similarly, when low-cost portble sensor measurements are included,
\begin{equation}
\textbf{X}_k = \textbf{H}_{X_k}\mathbf{Y} + \boldsymbol{\varepsilon}_{X_k}, \quad k > 0,
\end{equation}
where $\boldsymbol{\varepsilon}_{X_k}, k > 0,$ is normally distributed with mean $\mathbf{0}$ and covariance matrix $\sigma^2_{\boldsymbol{\varepsilon}_X}\textbf{I}$ and $\textbf{H}_{X_k}, k > 0,$ also contains one non-zero per row. Recall that $k$ denotes the step in the adaptive spatial design, $\textbf{X}_k$ is the data collected up to and including step $k$, $\textbf{H}_{X_k}$ is the incidence matrix corresponding to the low-cost sensors at step $k$, and $\boldsymbol{\varepsilon}_{X_k}$ are the corresponding measurement errors. Thus, this observation model is updated for each $k$ in the spatial design.

Now consider outputs from gridded proxy models $\mathbf{Q}$, such as CMAQ. The gridded outputs will span one or more BAUs, and hence we can write
\begin{equation}\label{eq:Z2}
\textbf{Q} = \textbf{H}_Q\mathbf{Y} + \boldsymbol{\varepsilon}_Q,
\end{equation}
where now $\textbf{H}_Q$ can contain more than 1 non-zero per row, and where each row of $\textbf{H}_Q$ sums to 1. The sum-to-one constraint ensures that we model each gridded model output as a process \emph{average}. On regular gridded domains, the non-zero elements of $\textbf{H}_Q$ are identical in value and equal to the reciprocal of the number of BAUs that are nested in each of the grid cells of the model output.

All the observations can be combined to yield the observation model at step $k$,
$$
\widetilde{\Zvec}_k = \textbf{H}_k\mathbf{Y} + \boldsymbol{\varepsilon}_k, \quad k \ge 0,
$$
where $\widetilde{\Zvec}_k \equiv (\Zvec', \mathbf{Q}', \mathbf{X}_k')'$, $\textbf{H}_k \equiv (\textbf{H}_Z', \textbf{H}_Q', \textbf{H}_{X_k}')'$, $\boldsymbol{\varepsilon}_k \equiv (\boldsymbol{\varepsilon}_Z', \boldsymbol{\varepsilon}_Q', \boldsymbol{\varepsilon}_{X_k}')',$ and where these definitions have obvious modifications for the case $k = 0$. Hence, for each $k$, $p(\widetilde{\Zvec}_k \mid  \mathbf{Y})$ is Gaussian as is $p(\mathbf{Y})$. Standard Gaussian identities can be used to obtain the conditional expectation $\hat\Yvec_k \equiv E(\mathbf{Y} \mid \widetilde\Zvec_k)$ and the conditional variance $\mathrm{var}(\mathbf{Y} \mid \widetilde\Zvec_k)$. Unknown parameters that appear in the model, namely $\sigma^2_\xi$ and those appearing in $\textbf{K}$, also need to be estimated; the process of iterating between maximum likelihood estimation of the parameters and prediction of $\mathbf{Y}$ is formalized in the expectation-maximization (EM) algorithm; full details are given in \citet{zammit:18}.

On convergence of the EM algorithm at step $k$, one has access to the prediction and, importantly, the prediction variance $\mathrm{var}(\mathbf{Y} \mid \widetilde\Zvec_k)$. The prediction variance $\mathrm{var}(Y(\svec) \mid \widetilde\Zvec_k)$ at a location $\svec$, as used in Equation \eqref{eq:utility}, is then just the prediction variance of the BAU containing $\svec$, that is, $\mathrm{var}(Y(\svec) \mid \widetilde\Zvec_k) = \mathrm{var}(Y_i \mid \widetilde\Zvec_k)$ for $\svec \in A_i$. As discussed in Section~\ref{sec:adaptsamp}, if repeated deployment and data collection is infeasible, the EM algorithm can be used to estimate the parameters just from $\Zvec$ and $\Qvec$, following which the adaptive algorithm can be run to obtain the entire set $\Sset^X$ without the need to collect additional data.

\subsection{Performance Evaluation}\label{sec:metrics}

We assess the efficiency of the adaptive sampling scheme by considering both the prediction accuracy and uncertainty quantification performance at a set of $n_p$ validation locations $\Sset^* \equiv \{\textbf{s}_1^*,\dots,\textbf{s}_{n_p}^*\}$. As metrics for solely assessing the prediction accuracy we used the mean absolute prediction error (MAPE) given by
$MAPE = \frac{1}{n_p}\sum_{p=1}^{n_p} |Y(\textbf{s}^*_p)-\hat Y(\textbf{s}^*_p)|,$
the root-mean-squared prediction error (RMSPE) given by
$$RMSPE = \sqrt{\frac{1}{n_p}\sum_{p=1}^{n_p} (Y(\textbf{s}^*_p)-\hat Y(\textbf{s}^*_p))^2},$$
and the mean prediction error (MPE) given by
$$MPE = \frac{1}{n_p}\sum_{p=1}^{n_p} (Y(\textbf{s}^*_p)-\hat Y(\textbf{s}^*_p)),$$
where $Y(\textbf{s}^*_p)$ and $\hat Y(\textbf{s}^*_p) = E(Y(\textbf{s}^*_p) \mid \widetilde\Zvec_{n_K})$ denote the realization of the underlying process at the validation location $\textbf{s}^*_p$ and the prediction obtained from the three different sources of data when adaptive sampling is complete, respectively. As a metric for assessing uncertainty quantification performance we used the continuous-ranked probability score (CRPS); see \citet{gneiting:07} for details on this proper scoring rule. These metrics are used to compare the adaptive sampling strategy of Section \ref{sec:adaptsamp} against a purely random sampling strategy in Section~\ref{sec:casestudies}.

\section{Case Studies}\label{sec:casestudies}

In this section we evaluate the impact of low-cost portable sensor network design on the quality of the predictions through two experimental case studies. In the first case, we examine the sensitivity of the prediction performance with respect to the signal-to-noise ratio (SNR) and the number of low-cost portable sensors to be deployed in an ideal setting where the underlying process is stationary and isotropic, and the domain is enclosed by a square. In the second case study we carry out an observing system simulation experiment, where the process is a simulation output from CMAQ in the Buffalo-Niagara region within the Erie and Niagara counties of western New York, USA. Here, the geometry of the domain is more complex and multiple risk factors are explicitly considered in the adaptive spatial sampling design.

The statistical analyses and adaptive spatial sampling  were conducted in \texttt{R} (v.3.5.1) \citep{R}. We used the \texttt{geosample} package (v.0.2.1) for  adaptive spatial sampling; the \texttt{FRK} package (v.0.2.2) for data fusion; the \texttt{verification} package for computing the CRPS; and the \texttt{rsm} package (v.2.10) for interaction effect analysis among the experimental design factors \citep{lenth:09}.

\subsection{Simulation Experiment}\label{sec:casestudies1}

The study area we considered is $D \equiv [0,1] \times [0,1]$. We assumed that the underlying process has a zero mean and an isotropic exponential covariance function given by $C_\thetab(\hvec) \equiv \textrm{cov}(Y(\svec), Y(\svec+{\bf h})\mid\thetab) = \sigma_Y^2 \exp(-||{\bf h}||/ \tau)$, where $\thetab \equiv (\sigma_Y^2, \tau)'$. Specifically, we assumed that the process has unit variance ($\sigma_Y^2$ = 1) and an e-folding length of 0.3 ($\tau = 0.3$). We obtained multiple realizations of the process on a 100 $\times$ 100 regular grid on $D$, and for each realization of the underlying process $\Yvec$, we generated two synthetic data sets of PM$_{2.5}$ concentrations, denoted as $\Zvec$ and $\Qvec$, respectively.

\begin{figure}[t!]
\centering
\includegraphics[scale=.15]{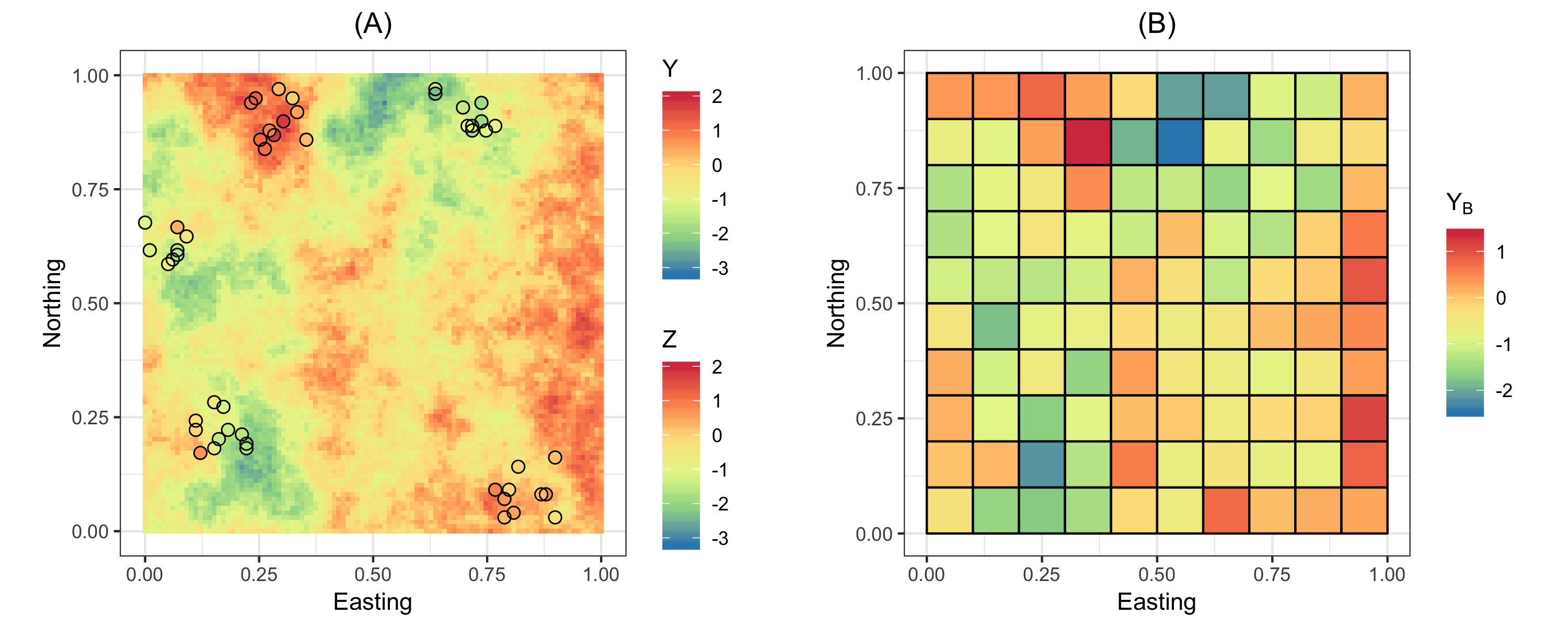}
\caption{(A) Simulated process on a $100\times 100$ grid on $D = [0,1] \times [0,1]$ and measurements $\Zvec$ at regulatory monitoring locations. (B) The process averaged on a coarser 10 $\times$ 10 grid, $\Yvec_B$.}
\label{fig:referenceValues}
\end{figure}

We assumed that the generated data are unbiased, but that they contain source-specific levels of measurement error. Specifically, regulatory fixed monitoring data $\Zvec$ were generated by taking the process values at $n_Z =$ 50 locations and adding measurement error $\epsilon_Z$ generated from a Gaussian distribution with zero mean and covariance matrix $\sigma_{\epsilon_Z}^2\Imat \equiv \frac{\sigma_Y^2}{SNR_Z}\Imat$. The signal-to-noise ratio of $\Zvec$ was set to 9 (SNR$_Z = 9$). The location of 50 regulatory fixed monitoring sites were selected in a non-regular form to mimic existing monitoring networks. Specifically, we selected 50 sites in five spatial clusters, as shown in Figure \ref{fig:referenceValues}(A).

 To generate the gridded data, we first averaged the process values on a coarser 10 $\times$ 10 grid on $D$, thus resulting in a vector of process averages which we denote as $\Yvec_B$ (i.e., $\Yvec_B \equiv \Hmat_Q\Yvec$). The data $\Qvec$ were then obtained by adding on $\epsilonb_Q$ to $\Yvec_B$ as in Equation \eqref{eq:Z2}, where, for simplicity, $\epsilonb_Q$ was assumed to be uncorrelated, have zero mean, and an SNR of 1 (SNR$_Q = 1$). A single realization of the underlying process ${\bf Y}$ over the study region is shown in Figure~\ref{fig:referenceValues}(A) as a background surface and the ground observations at 50 monitoring stations are denoted as circle symbols on top, while the same process averaged under the coarser grid is shown in Figure~\ref{fig:referenceValues}(B). Once the new sampling locations were selected we generated the measurements $\Xvec$ in a similar manner to the data $\Zvec$, that is, by taking the process values at the selected locations and adding on measurement error $\epsilonb_X$ with zero mean and covariance matrix $\sigma_{\epsilon_X}^2\Imat = \frac{\sigma_Y^2}{SNR_X}\Imat$. Here, we considered a range of SNR values to assess the effect of data quality of low-cost sensor measurements: a larger SNR (SNR$_X$ = 9) for data with a small measurement-error variance like the measurements obtained from regulatory instruments, and a smaller SNR (SNR$_X$ = 1) for less accurate measurements from low-cost portable sensors. 

\begin{figure}[t!]
	\centering 
	\includegraphics[scale=.15]{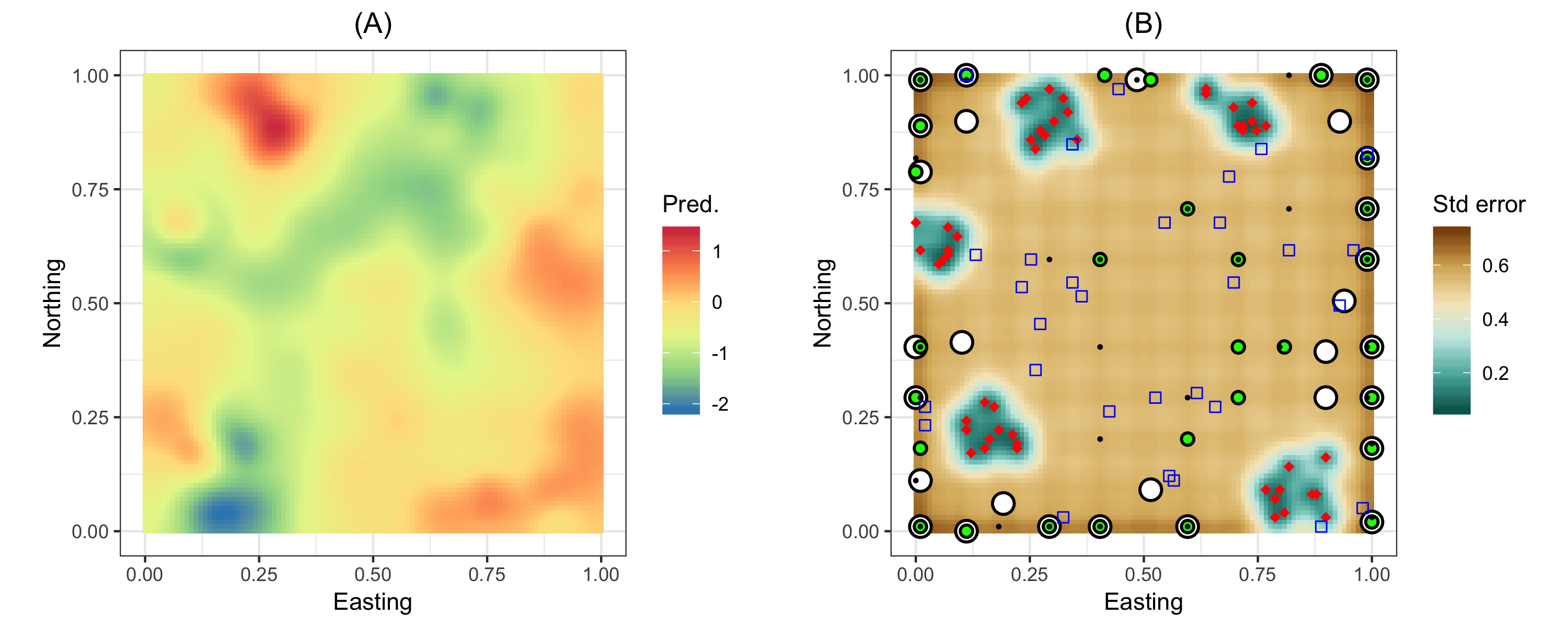}
	\caption{(A) FRK prediction obtained from $\Zvec$ and $\Qvec$ shown in Figure~\ref{fig:referenceValues}. (B) Associated FRK prediction standard error . In panel (B) we also show the sampling locations for both the regulatory and low-cost sensors. The red circles in regions of low FRK prediction standard error denote the 50 regulatory monitoring stations. The black circles with different sizes and filling denote the $n_X$ = 30 sampling locations selected through the adaptive sampling with different batch sizes. The largest circles with white filling denote the sampling locations with the batch size $b$ = 30, the medium size circles with green filling denote the sampling locations with the batch size $b$ = 15, and the smallest black dots denote the sampling locations with $b$ = 3. The locations selected from random sampling are shown as blue square symbols.}
	\label{fig:FRKwith3data_simulation}
\end{figure}

We considered two sampling strategies --- adaptive spatial sampling with a batch size  $b \in \{3, 15, 30\}$, and random spatial sampling, to select the $n_X$ locations. The sampling locations were selected at random for the latter approach, whereas they were selected based on FRK prediction standard error surfaces and the minimum spacing constraint of $\delta_d = 0.1$ for the adaptive spatial sampling approach. The prediction and prediction standard error surface based on only $\Zvec$ and $\Qvec$ (i.e., at step $k = 0$) from the realization of Figure~\ref{fig:referenceValues} are shown in Figures \ref{fig:FRKwith3data_simulation}(A) and \ref{fig:FRKwith3data_simulation}(B), respectively. In Figure \ref{fig:FRKwith3data_simulation}(B), we also show the 50 sites of the regulatory stations (denoted by red circles) and the additional $n_X = 30$ monitoring sites selected for portable sensors using both random sampling (blue squares) and adaptive sampling with different batch sizes $b \in \{3, 15, 30\}$. 

Our results showed that $b =$ 3 (10 \% of the total sample size) yielded the smallest value of utility function of Equation (\ref{eq:utility}) when averaged over $D$ (the lowest overall uncertainty) among the three batch sizes considered. We further investigated the effect of the batch size on the overall prediction accuracy and uncertainty for varying $n_X$ and $b$; we summarize the results in Appendix A. As expected, the larger $n_X$ is, and the smaller $b$ is, the better the prediction performance.  We note that the selected sampling sites for low-cost portable sensors concentrate in regions of high uncertainty which, in this example, occur at the border of $D$ regardless of the batch sizes chosen. Modifications to the utility that avoid this issue are discussed in \citet[]{krause:08}. When other constraints, such as spatially-resolved risk factors, are considered in the utility function, this effect will generally disappear; see Section \ref{sec:realistic}. 

The results above are based on a single process realization. We obtain more general results by evaluating the prediction performance using the metrics introduced in Section \ref{sec:metrics} under a variety of settings and across several realizations. We generated a set of 100 Monte Carlo (MC) simulations of the underlying process, from which each pairwise combination of low-cost portable sampling factors were considered, for $n_X \in \{10, 30, 50, 70\}$ and SNR$_X \in \{1,3,5,7,9\}$. The 100 MC simulations were conducted to account for the potential influence that the underlying process has on the prediction quality as well as the adaptive sampling results. We generated $\Zvec$ and $\Qvec$ for each of the 100 process simulations, from which we obtained prediction standard error surfaces using FRK, and determined $\Sset^{X^{(1)}}$ based on Equation \eqref{eq:optim} and Equation \eqref{eq:utility} with $T(\cdot) = 0$ for the adaptive sampling approach. For this experiment we fixed $b = 1$ since the results of Appendix A show that the smaller $b$ is, the better the prediction performance.  For each simulation, FRK was re-run using $\widetilde\Zvec_1 = [\Zvec', \Qvec', X^{(1)}]'$ and adaptive sampling was applied again to select $\Sset^{X^{(2)}}$.This process was repeated until all $n_X$ samples were selected.  Once these locations were selected, 30 random realisations of $\Xvec$ were generated, and for each of these realisations the diagnostics outlined in Section~\ref{sec:metrics} were computed. The end result of this experiment was a $100 \times 30$ table for each diagnostic and for every combination of $n_X$ and SNR$_X$. These tables were also obtained for the random sampling case, where the $n_X$ samples were randomly placed in $D$ and additional 30 MC simulations were conducted for each  SNR$_X$. For each case, FRK was used to obtain the prediction and prediction-error surfaces.

\begin{figure}[t!]
\centering
\includegraphics[scale=.6]{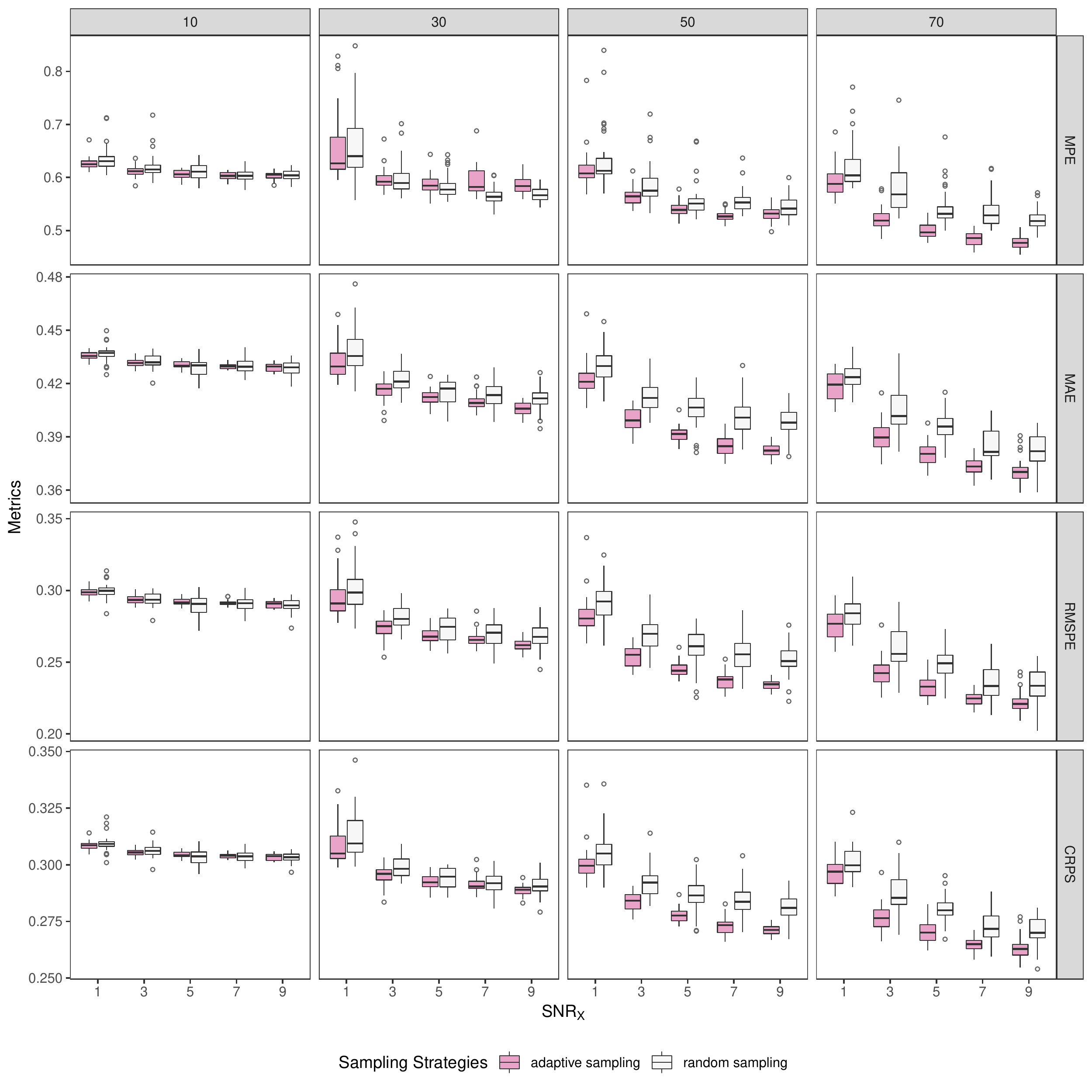}
\caption{The effects of size and quality of low-cost sensor data on prediction accuracy for a single realization of $Y(\cdot)$. In each facet the sampling effect is assessed over five different levels of SNR$_X$ using box plots across 30 MC simulations. The box plot in the left (shaded) corresponds to adaptive sampling and the right box (unshaded) to random sampling.}
\label{fig:MEeffect_aYsim}
\end{figure}

In Figure \ref{fig:MEeffect_aYsim} we show the results from a single realisation of the process (corresponding to one row in our $100 \times 30$ tables). From the first column ($n_X$ = 10) we see that the sampling strategies have no effect on the prediction quality, regardless of SNR$_X$, when a small number of low-cost portable sensors are deployed. In contrast, when a large number of sensors are deployed ($n_X$ = 70) as shown in the fourth column of Figure \ref{fig:MEeffect_aYsim}, the difference between the two strategies is substantial across all four metrics. It is also noticeable that the relative benefits of using adaptive sampling versus random sampling increase when the measurements are more reliable (SNR$_X$ = 9).  Therefore, as expected, an increased number of low cost sensors will result in a decrease of the prediction error regardless of the chosen sampling strategy; however adaptive sampling tends to be more effective than random sampling and especially so when the uncertainty of data from the low-cost portable sensors is relatively low. 

\begin{figure}[t!]
\centering
\includegraphics[scale=.8]{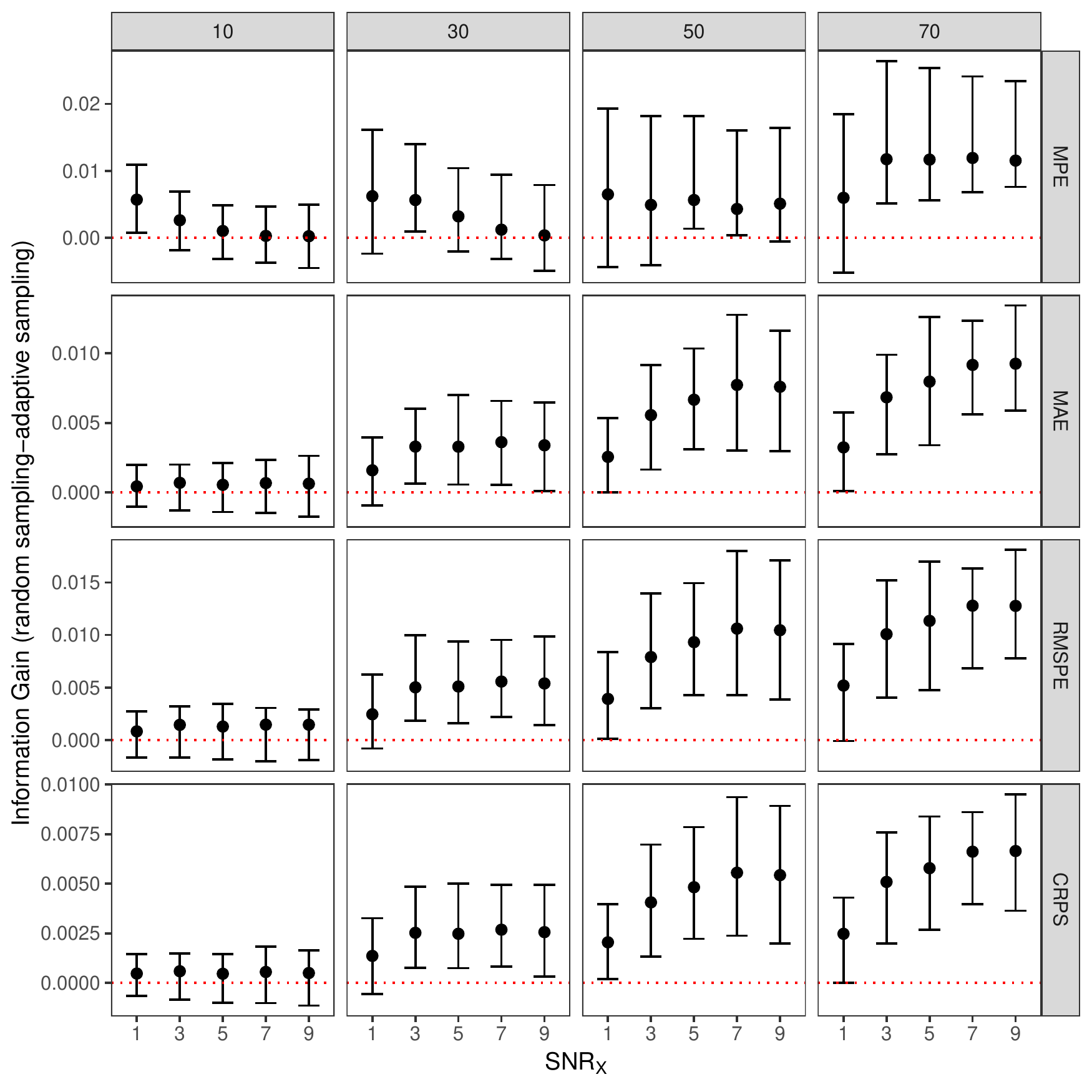}
\caption{The mean differences (random sampling $-$ adaptive sampling) of metrics associated with two sampling strategies, across 100 $\times$ 30 simulations, shown here by $n_X$ and $SNR_X$. The dots denote the mean of differences in metrics and the upper and lower bars represent the inter-quartile range of the differences.}
\label{fig:MEeffect_allSims}
\end{figure}

Recall that Figure \ref{fig:MEeffect_aYsim} is from just one realization of the underlying process. To assess the sensitivity of our findings with respect to all other realizations, we compared the magnitude of performance improvement gained from adaptive sampling versus random sampling by diagnostic. Specifically, we calculated the averaged differences of the metrics associated with each sampling strategy (random sampling $-$ adaptive sampling) across all realizations. When two sampling strategies give similar results, the difference should be close to zero. The results from this analysis are presented in Figure {\ref{fig:MEeffect_allSims}} where the summary statistics of the averaged differences are presented in a combination of metrics (rows) and sample sizes (columns). Overall, adaptive sampling outperforms random sampling given that all the mean differences are greater than or equal to zero (on or above the dotted lines) indicating that the prediction error from adaptive sampling was less than, or equal to, that of random sampling. The effect of the sampling strategy is clear for a large sample size ($n_X$ = 70), as expected from the analysis of a single process realization. For example, the metrics in the fourth column of Figure {\ref{fig:MEeffect_allSims}} show that the mean differences are above zero regardless of SNR$_X$ values. On the other hand, it is also clear that the quality of low-cost sensors (SNR$_X$) has little to no effect on the performance gain associated with the sampling strategy when the sample size is small ($n_X$ = 10) (see the first column of Figure {\ref{fig:MEeffect_allSims}}). Even when the sample size is sufficiently large ($n_X$ = 70), the effect that the quality of measurements (increased values of SNR$_X$) has on the performance improvement from the adaptive sampling is not linear. After a sharp increase of performance gain with respect to the increased SNR$_X$ (from 1 to 3, for example), the performance improvement is incremental despite the the higher values of SNR$_X$.

\begin{figure}[t!]
\centering
\includegraphics[scale=.65]{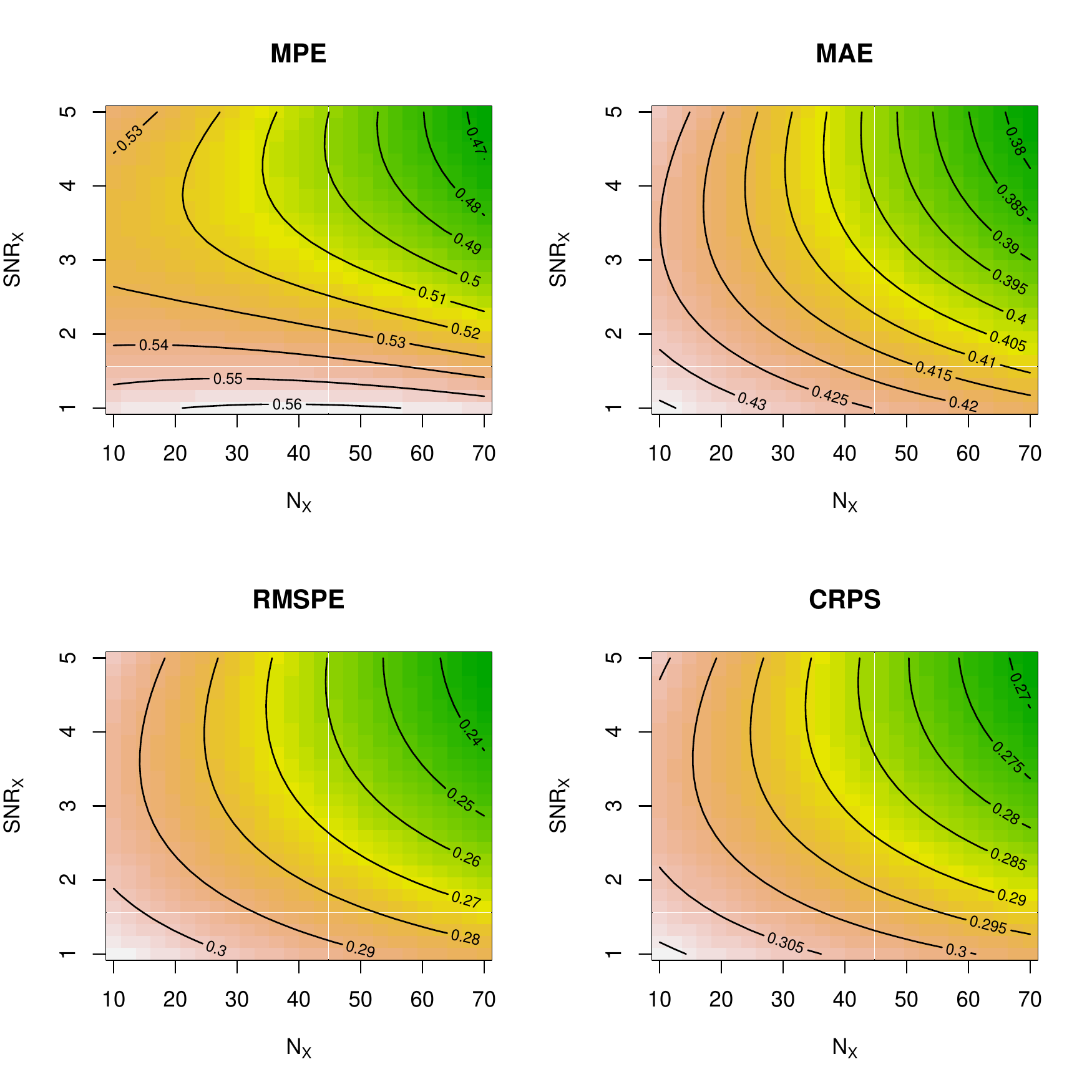}
\caption{Interaction effect of the sample size ($n_X$) and the data quality (SNR$_X$) of low-cost sensor measurements on the four metrics.}
\label{fig:Interactioneffects}
\end{figure}

Finally, we assessed the interaction effect between the sample size ($n_X$) and data quality of low-cost sensor measurements (SNR$_X$) on the performance of adaptive sampling in terms of the four metrics through response-surface analysis. In Figure \ref{fig:Interactioneffects} we show the response surfaces of the four metrics as contour-surface plots. Clearly as the sensor count $n_X$ increases and the data quality improves (SNR$_X$ increases), all four metrics improve. Interestingly, the functions are reasonably similar in shape, and one could also potentially use these surfaces to predict the potential performance improvement that can be obtained from increasing $n_X$ and SNR$_X$ in a cost-effect analysis.

\subsection{Observing system simulation experiment of PM2.5 concentrations in the Erie and Niagara counties}\label{sec:realistic}

\begin{figure}[t!]
\centering
\includegraphics[scale=.6]{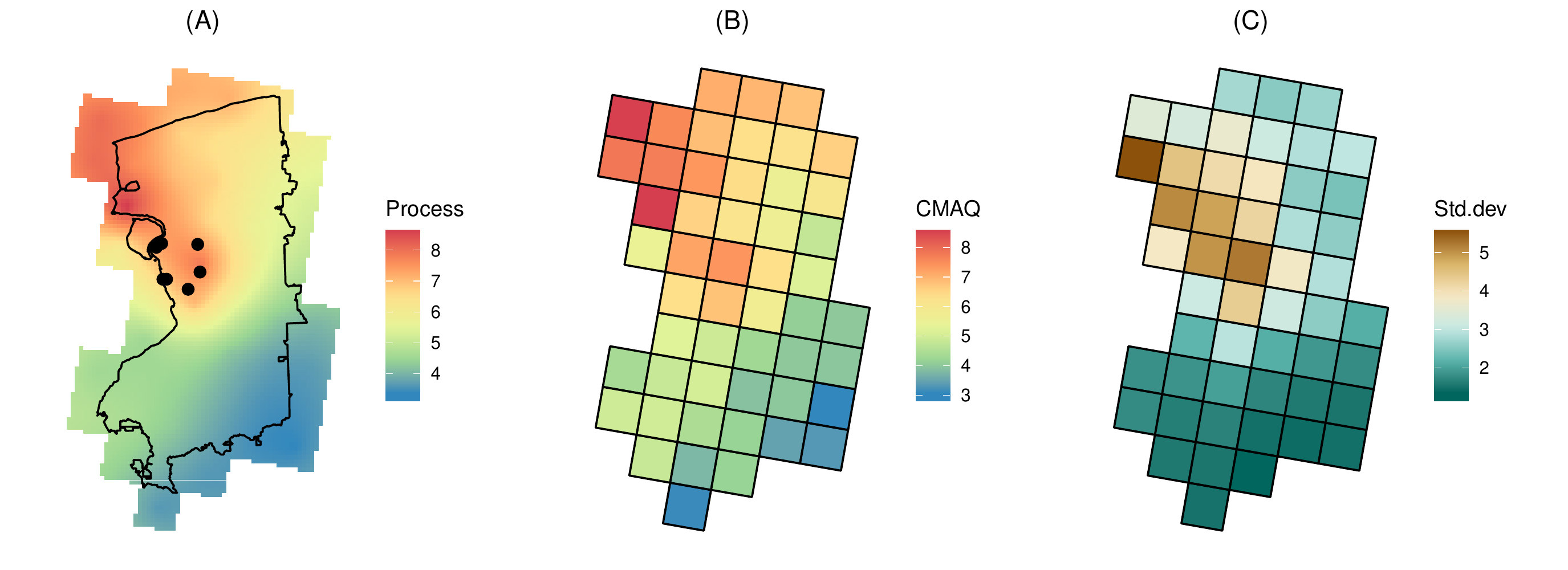}
\caption{(A) Simulated process of PM$_{2.5}$ concentrations for Erie/Niagara counties and the locations of the state ambient air monitoring stations (black dots). (B) Simulated PM$_{2.5}$ concentrations that represent CMAQ output. (C) Uncertainty associated with the simulated CMAQ outputs.}
\label{fig:referenceValues_EN}
\end{figure}

In this experiment we simulated process values on a regular $1 \times 1$ km$^2$ grid over the Erie and Niagara counties in Western New York, USA. However, unlike the unconditional simulation used in Section \ref{sec:casestudies1}, here the process values were simulated conditional on the daily PM2.5 mass concentrations obtained from a CMAQ model simulation on 09/02/2011. Specifically, the CMAQ output was first made point-referenced, by concentrating the data at the geometric centroids of the CMAQ grid cells. Following this, a PM2.5 concentration surface on the $1\times 1$ km$^2$ grid was generated using ordinary kriging with the point-referenced CMAQ output as data. A constant mean and an exponential covariance function $C_\thetab(\hvec) = \sigma_Y^2 \exp(-||{\bf h}||/ \tau)$, with $\sigma_Y^2=$ 0.76 and $\tau =$ 35.8 km, were used. A conditional realization of the process, which we take as our true process, is shown in Figure \ref{fig:referenceValues_EN}(A). 

We treated the $1 \times 1$ km$^2$ grid cells as BAUs, and hence every point measurement at the air quality monitoring stations and the low-cost portable monitoring sites were all attributed to a grid cell. A total of seven existing ambient air monitoring stations  $\{\svec_{Z_i}: i=1, \ldots, 7\}$ were operated in the study area during 2011, which are shown in Figure \ref{fig:referenceValues_EN}(A). We generated the ground PM2.5 measurements at the seven monitoring stations by taking the process values at the collocated locations, and adding measurement error $\epsilon_Z$ with zero mean and covariance matrix $\sigma_{\epsilonb_Z}^2\Imat \equiv \frac{\sigma_Y^2}{SNR_Z}\Imat$ with SNR$_Z$ = 7. Simulated areal data representing CMAQ measurements were generated by averaging the process values over a 12 $\times$ 12 km$^2$ regular grid $\{B_i: i=1, \ldots, 51\}$ and adding uncorrelated measurement error $\epsilonb_Q$ with zero mean and spatially-varying variance. Given that CMAQ is a deterministic model based on atmospheric science and air quality modeling techniques that include multiple emission sources and atmospheric chemistry, the uncertainty associated with the simulation outputs is not readily available. 
We instead use the temporal variability of the process to represent our uncertainty though $\epsilonb_Q$. Specifically, we summarize the variability present in 15 consecutive days of daily CMAQ values --- seven days prior to and seven days after 09/02/2011 --- at each pixel in order to obtain our measure of uncertainty. The daily CMAQ outcomes and corresponding temporal variabilities, in terms of the empirical standard deviation, are presented in Figures \ref{fig:referenceValues_EN}(B) and \ref{fig:referenceValues_EN}(C), respectively.

Summary statistics for the simulated process values and the simulated concentrations obtained at the seven air quality monitoring stations and on the CMAQ grid are summarized in Table \ref{tbl:summaryStat}. The mean values of the process over the entire domain and the air monitoring stations are 5.37 and 7.16, respectively, with the range of the latter being narrower (6.25 and 7.65) than the range of the simulated values (3.26 and 8.52). This is somewhat expected since the regulatory monitoring stations are concentrated in a small part of the domain. On the other hand, the mean and the standard deviation of the CMAQ PM$_{2.5}$ concentrations are 5.42 and 1.46, respectively, which are very similar to that of the simulated process (5.37 and 1.41, respectively). 

\begin{table}[t!]
\centering
\caption{Summary statistics of process values and simulated data}
\begin{tabular}{lrrrrrrrr}
  \hline
 & Min. & 1st Q. & Median & Mean & 3rd Q. & Max. & SD & N\\ 
  \hline
Process Values &  3.26  &  4.08  &  5.26 &   5.37  &  6.66  &  8.52   & 1.41 &  4990 \\ 
Regulatory Stations  & 6.25 &   6.68 &   7.41  &  7.16  &  7.53  &  7.65  &  0.51  &  7 \\ 
CMAQ Outputs  & 2.95   & 4.09  &  5.36  &  5.42  &  6.56  &  8.45  &  1.46  & 51\\
   \hline
\label{tbl:summaryStat}
\end{tabular}
\end{table}

\begin{figure}[t!]
\centering
\includegraphics[scale=.65]{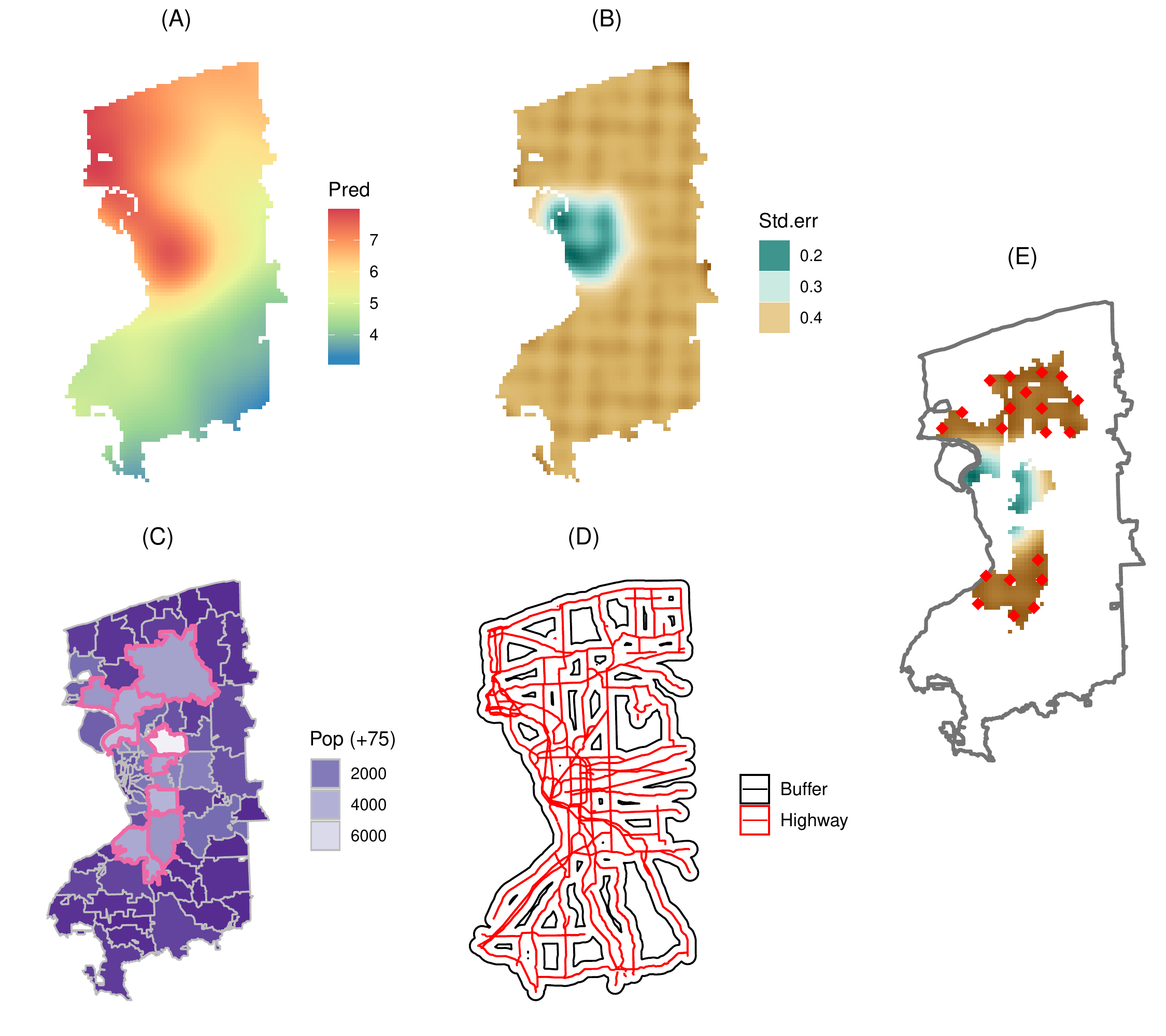}
\caption{(A) Prediction from FRK when using the simulated data at the regulatory monitoring station locations and the simulated CMAQ output. (B) The associated prediction standard error surface. (C) Zip code units with those in the top-ten percentile with respect to the proportion of individuals aged 70 years and above outlined in pink-colored thick line. (D) Major highways (red) and a 2 km buffer zone (black). (E) The prediction standard error surface of (B) at the pre-determined high-risk locations. The red dots indicate the chosen locations following adaptive sampling.}
\label{fig:riskFactors}
\end{figure}

The first step of our adaptive sampling algorithm requires us to obtain the prediction uncertainty through FRK using $\Zvec$ and $\Qvec$. The  predictions and prediction standard errors from running FRK on these data are shown in Figures \ref{fig:riskFactors}(A) and \ref{fig:riskFactors}(B), respectively. The effect of the ground observations from the seven existing monitoring stations on the prediction uncertainty is clear: The prediction error variance is relatively high and homogeneous outside the central western part of the study region where the seven EPA monitoring sites are located.

For adaptive sampling, we incorporated other risk factors in the site selection process to represent spatially-varying demands. The risk factors included a set of zip code units that have high concentrations of the elderly population, shown  in Figure \ref{fig:riskFactors}(C), and the proximity to major highways (2 km buffer zones from highways), shown in Figure \ref{fig:riskFactors}(D). Specifically, we selected zip code units that are in the top-ten percentile with respect to the proportion of individuals aged 75 years and above. Let $R_1(\cdot) = 1$ inside the high-risk zip code units and zero otherwise, and let $R_2(\cdot) = 1$ in regions with 2 km or less away from the highways and zero otherwise. We let $T(\cdot) = R_1(\cdot) \times R_2(\cdot)$ and set $\lambda = 1.0$ in Equation~\eqref{eq:utility} to make it highly unlikely that portable sensors are placed outside the risky areas. Spatial locations where $T(\cdot) > 0$ are depicted as colored grid cells in Figure \ref{fig:riskFactors}(E) under the red dots. The colors of the grid cells represent the prediction uncertainty within the high-risk zones.

The adaptive spatial sampling was conducted with $n_X = $ 20 under the constraint of $\delta_d = 3$ km. Our adaptive sampling procedure with $b = 1$ identified 20 sites for low-cost sensor sampling; these are denoted as red dots in Figure \ref{fig:riskFactors}(E). At each step in the algorithm a new low-cost sensor measurement was generated by taking the process value at the newly selected  site and adding measurement error with SNR$_X = 4$. 

\begin{figure}[t!]
\centering
\includegraphics[scale=.7]{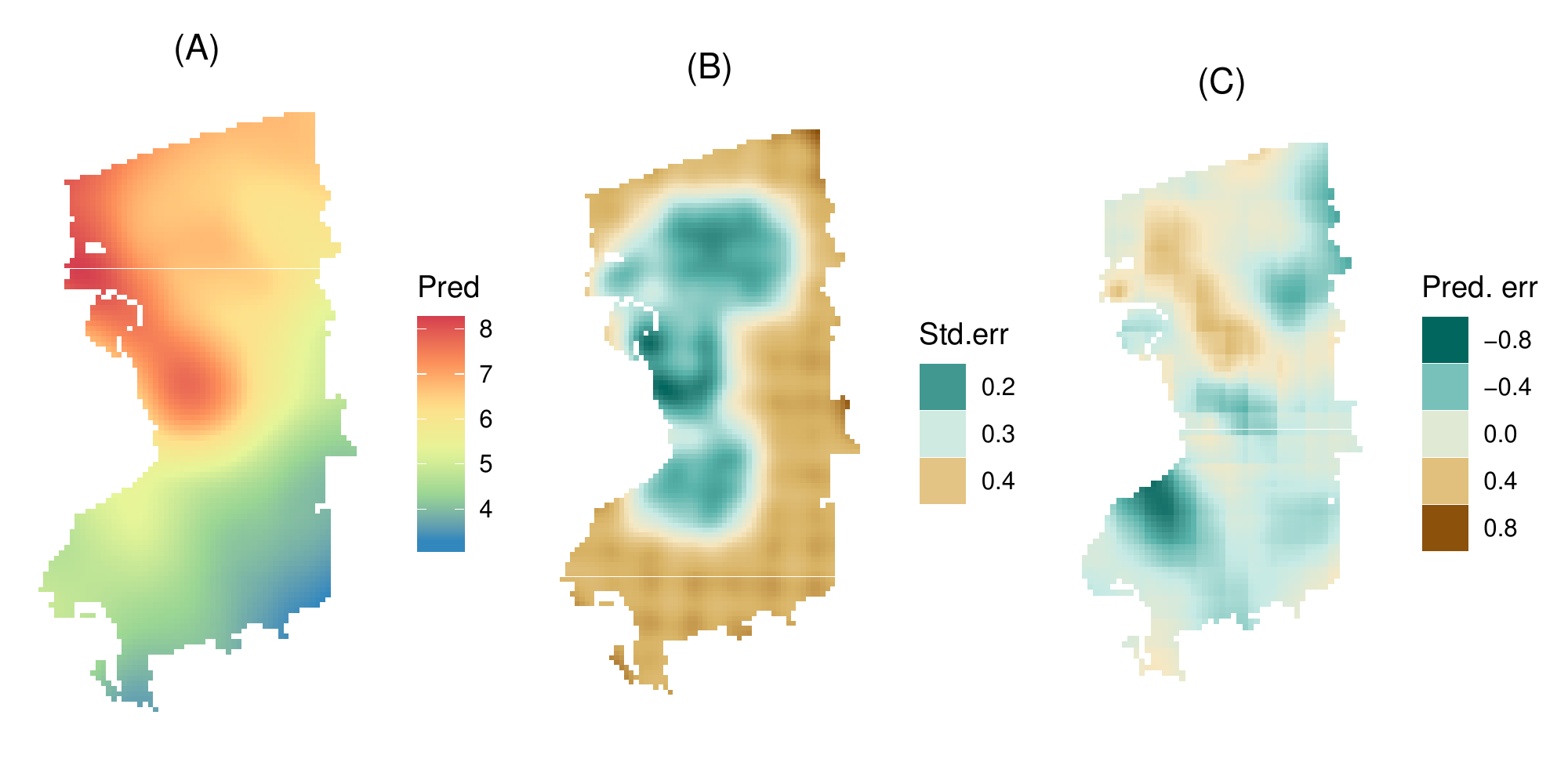}
\caption{(A) FRK prediction when using all three data sources. (B) The associated prediction standard error. (C) The prediction errors (Figure 8(A) $-$ Figure 6(A)).}
\label{fig:FRK_3data}
\end{figure}

The resulting prediction and prediction standard error surfaces obtained by running FRK on all the three data sets are illustrated in Figures \ref{fig:FRK_3data}(A) and \ref{fig:FRK_3data}(B), respectively. The predictions lie between 3.19 and 7.86, a range that is slightly narrower than that of the process (3.26 to 8.52). Both the mean (5.57) and and the standard deviation (1.26) of the predictions corroborate those of the process  (See Table \ref{tbl:summaryStat}). Meanwhile the difference between the predictions with and without the low-cost sensor measurements does not appear to be substantial; compare Figure \ref{fig:riskFactors}(A) to Figure \ref{fig:FRK_3data}(A). This is likely due to the fact that the underlying process is relatively smooth. However, the effect of new sample data is clear in terms of the prediction standard error --- we now have low prediction standard error (around 0.2 and 0.3) in larger regions including areas where the portable sensors are placed; compare Figure \ref{fig:FRK_3data}(B) to Figure \ref{fig:riskFactors}(B). We further assessed the accuracy of the prediction with three data sources by calculating the prediction error at each prediction location. The prediction error shown in Figure \ref{fig:FRK_3data}(C) suggests an overall good agreement between the prediction and the underlying process (mean of -0.09 and SD of 0.17).  

To better understand the contribution of the low-cost portable sensor measurements on prediction accuracy while accounting for the uncertainty associated with measurement error in the low-cost portable sensor measurements, we ran the above experiment using 100 different simulations of measurement errors associated with $\Xvec$. Here we fixed the sampling locations obtained from the initial adaptive sampling strategy, as shown in Figure \ref{fig:riskFactors}(E). For each realization, we assessed the prediction accuracy using the four metrics discussed in Section \ref{sec:metrics}, and compared them to those based on only the two data sets of $\Zvec$ and $\Qvec$. As shown in Table \ref{tbl:MAPE summary}, FRK predictions with three data sets including the low-cost sensor measurements largely improve the prediction quality, but not always. For example, the 0.75 quantile of MPEs obtained from 100 MC runs is greater than the MPE obtained with just the two data sets. On the other hand, the MAE and the RMSPE improved considerably after additional data was included for prediction.

\begin{table}[t!]
\caption{The effect of low-cost portable sensor data ($\Xvec$) on spatial prediction accuracy as obtained through MC simulation. For definitions of the metrics used see Section~\ref{sec:metrics}.}
\begin{center}
\begin{tabular}{l|c|ccccc}
Metric &  Prediction without $\Xvec$ & \multicolumn{5}{c}{Prediction with 100 MC simulations of $\Xvec$}  \\ 
 &  & Min & Q$_{.25}$ & Q$_{.50}$ & Q$_{.75}$ & Max \\ \hline
MAE & 0.19 & 0.12 & 0.14 & 0.14 & 0.15 & 0.17 \\ 
MPE &0.43  & 0.40 & 0.42 & 0.43 & 0.51 & 0.65 \\ 
RMSPE &0.06 &0.02 & 0.02 & 0.02 & 0.02 & 0.03 \\ 
CRPS &  0.15& 0.13 & 0.14 & 0.15 & 0.15 & 0.19 \\ 
\end{tabular}
\end{center}
\label{tbl:MAPE summary}
\end{table}%

\section{Discussion}

Recent advances in sensing technologies have enabled researchers and communities to collect additional data for environmental monitoring, but the quality of these additional data and their spatial resolutions are not yet comparable with those of existing regulatory instrument measurements. There is a rich literature on spatial  sampling design strategies, but they do not necessarily account for change-of-support problems nor the heterogeneous data quality among the multiple data sources. In this article we combined an adaptive spatial sampling design with a general spatial data fusion approach to determine optimal sampling sites for a fixed number of low-cost portable sensor deployment.

The main contributions of this article are two-fold. We explicitly accounted for the quality of an emerging data source, low-cost portable air sensor measurements, for optimal site selection in terms of quantifying the design criteria by applying a state-of-the-art statistical spatial data fusion approach, namely, fixed rank kriging (FRK). The proposed framework based on FRK is flexible in that it accommodates the support differences among the spatial data and the different quality of each data source, while being capable of handling potentially large data. In the optimal sampling site selection algorithm for low-cost portable sensors, we applied an adaptive sampling approach under the explicit consideration of uncertainty, and account for multiple risk factors that determine the realistic needs of policy makers.

Through two case studies we examined the impact of the sampling strategies on the prediction accuracy using two synthetic data sets. In the first case study we assessed the effect of the total number of sensors and the quality of the sensors on the prediction quality, using an experimental design with simulated data in an ideal but rather simplified setting, where the underlying process was stationary and isotropic and the domain was enclosed by a square.  The prediction quality was quantified using four metrics. The results show that adaptive sampling is more effective than random sampling in general, but that its effects are more pronounced when a large number of sensors are deployed. In addition, the quality of low-cost portable sensor measurements positively affects the prediction accuracy but only tangibly so when the sample size is sufficiently large. Specifically, the relative performance gain of adaptive sampling may not be evident if the sample size is too small. Similarly, we should not expect that prediction performance increases linearly with data quality (the quality of low-cost sensors). These results suggest that the proposed design framework has the potential to be useful for decision making in a resource-poor conditions where the choice between quantity and quality has to be made.

It should be noted that our experimental design did not address the question of how many additional data points are sufficient, or how much measurement error with low-cost portable sensors is acceptable, to satisfy a pre-determined criterion. Rather, we focused on the effect of a design strategy for a  given number of sensors of known quality. In the second case study we demonstrated how other conditions, such as risk factors, can be incorporated in the adaptive sampling, which has practical implications in real-world applications. However, the results confirmed our findings from the first case study in that the low-cost sensor measurements do not universally guarantee an improvement in prediction performance. Guarantees can only be attained when a large number of portable sensors are used. 

Our experiments used synthetic data, but we are motivated by real-world policies. Our findings suggest that additional low-cost sensor measurements have the potential to significantly improve our understanding of individual air pollution exposure, as they enable investigators to quantify and characterize exposure gradients within urban areas that can be used in epidemiological studies. Additionally, fine scale air quality information is likely to assist regulators with policy making and planning. For example, the Village Green project (\url{https://www.epa.gov/air-research/village-green-project}) supported by US Environmental Protection Agency aims to provide the public and communities with the information on real time air quality using low-cost air sensors. Importantly, our results raise questions as to when and if portable sensors should be used, as we have seen that their use may not be as efficient as anticipated unless a sufficient number are deployed or the quality of sensors meets a certain standard. For example, adding one or two portable sensors which are of relatively low-quality is wasteful if the underlying concentration field is spatially smooth. These concerns are exacerbated by the fact that low-cost sensors are frequently sensitive to environmental conditions and require field-calibration \citep{zimmerman:18}.

The proposed sampling design aims to improve the quality of long-term predictions of air pollutants whose small-scale (within-city) spatial variation is so substantial that it cannot be captured by existing monitoring networks. By design, however, monitoring based on the proposed method can be temporally limited. To minimize the potential bias, the proposed sampling design can be undertaken multiple times over the study period to account for seasonal variations in air pollutant concentrations or to avoid the influence of abnormal weather or events, such as holidays or wildland fires, as suggested by similar studies \citep{hoek:02,henderson:07,madsen:07}. Another important limitation includes the use of CMAQ as a proxy variable. CMAQ has potentials to fill the spatial and temporal gaps in ground observations and enables investigators to estimate emission source specific air pollutant concentrations, but its computational cost is high and using it for real-time estimation is impossible. Alternatively, AOD or diffusion model outputs can be considered as a proxy variable for regional air quality data. \citep{pu:19}.

There are several avenues for future work. First, in our two case studies we have assumed that the spatial structure of the underlying process is known, specifically that it has a constant mean and an exponential covariance function. However, environmental processes often change over time and their spatial patterns may be heterogeneous. The proposed sampling design for long-term exposure assessment of outdoor air pollutant can be extended to update the mean and covariance to accommodate the temporal variation in air pollution for short-term exposure studies. Second, we assumed that the measurement error standard-deviation is known for all data sources, but this is unrealistic in real-world applications. Finally, the batch size $b$ used in adaptive spatial sampling was seen to have an  influence on the optimal sampling results. Further investigation on the influence of the batch size $b$ when optimizing the utility function is thus warranted.

\section{Conclusions}
Emerging new technologies bring both opportunities and challenges to scientific communities, including the atmospheric science and environmental exposure science ones. Although the affordability and portability of low-cost air monitoring sensors enables researchers and communities to overcome sparse data issues by collecting additional air quality measurements at times and locations of interest, concerns about the accuracy of these measurements remain a challenge. Moreover, the needs for deploying a large number of sensors across a small geographic area spatially vary as the air pollution level over areas with lower population densities and less traffic congestion would be significantly different from that of large urban areas with high population densities and significant traffic congestion. To achieve the maximum utility of low-cost portable sensors, our study suggests the use of efficient spatial sampling strategies for sensor deployments and the integration of measurements from low-cost portable sensors with these from existing data sources, such as point measurements from regulatory monitoring stations or gridded proxy data from atmospheric models. Specifically, we propose an adaptive spatial data sampling design approach within a flexible spatial data fusion framework, namely fixed rank kriging, for efficient low-cost portable sensor data collection. Based on a multi-factorial simulation experiment, we conclude that low-cost portable sensors are only likely to be beneficial if they are sufficient in number and quality. Our study was presented in a spatial setting, but the approach could be extended to spatio-temporal settings by accounting for the dynamically evolving process underlying the observations. Beyond air pollution, we anticipate that the proposed design framework is readily applicable to other environmental science problems that involve combining heterogeneous spatial data sets with different supports and measurement-error characteristics.

\section*{Acknowledgments}
EHY acknowledges the support of the Center for Computational Research and the Research and Education in Energy, Environment and Water (RENEW) seed project funding at the University at Buffalo. AZM was supported by an Australian Research Council (ARC) Discovery Early Career Research Award, DE180100203.

\bibliography{ijgis}
\renewcommand{\thefigure}{A\arabic{figure}}

\setcounter{figure}{0}
\appendix

\titleformat{\section}
{\normalfont\Large\bfseries}{Appendix~\thesection}{1em}{}

\newpage

\section{The effects of batch size on prediction variance and accuracy}

\begin{figure}[htt!]
\centering
\includegraphics[scale=.7]{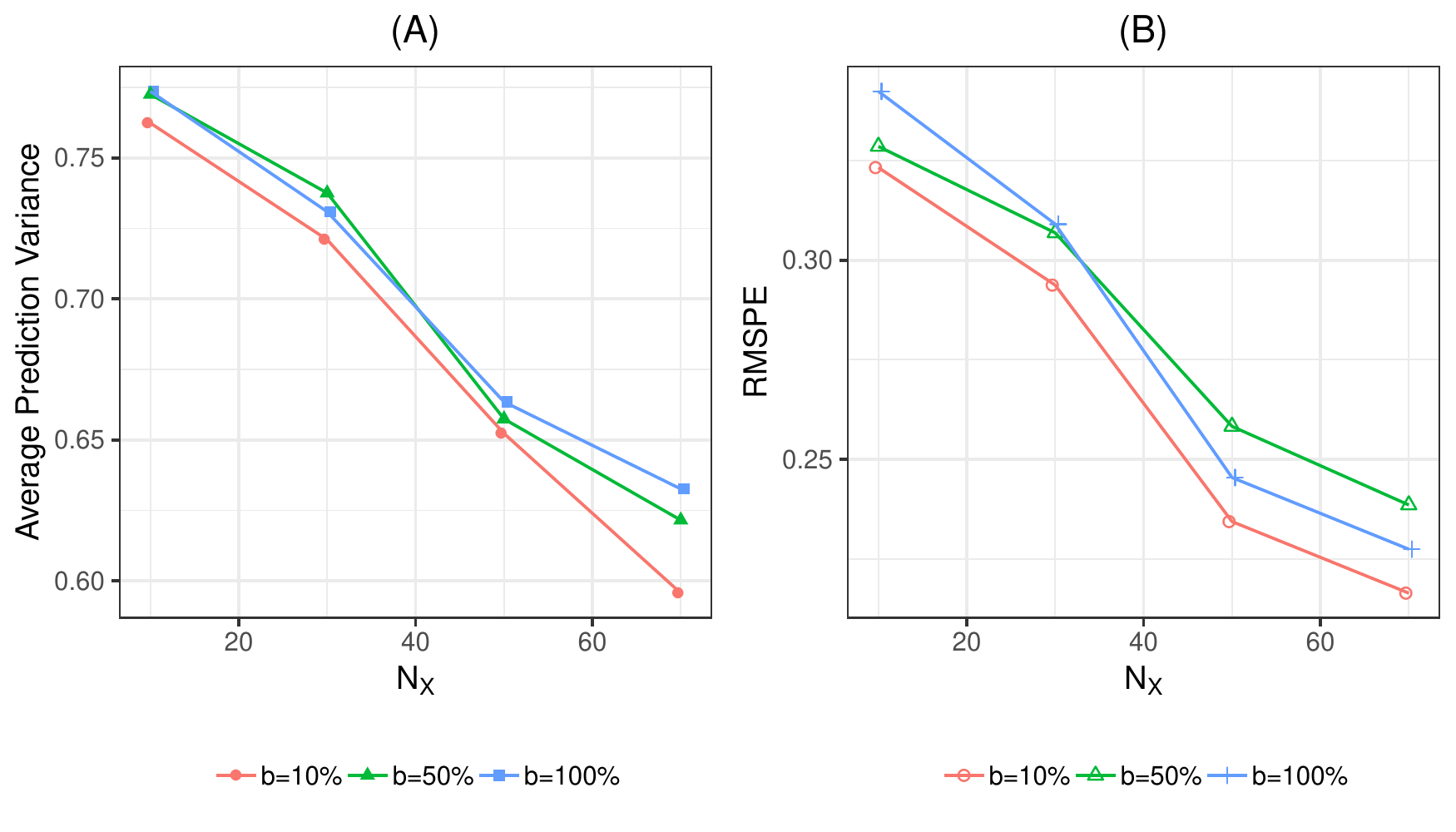}
\caption{Batch adaptive sampling with batch sizes $b = $3, 15, and 30, with the initial sample size $n_Z =$ 50 as shown in Figure \ref{fig:referenceValues}. For each $n_X$, the red circles, green triangles, and blue squares, denote the batch sizes 3 (10 \%), 15 (50 \%), and 30 (100\%) of $n_X$ for (A)  the mean of prediction error variance and  (B) the root-mean-squared prediction error (RMSPE), respectively.}
\label{fig:batchsize}
\end{figure}

Based on the FRK prediction standard error surface obtained from $\bf Z$ and $\bf Q$ (see Figure \ref{fig:FRKwith3data_simulation}), we selected four sets of sampling locations for low-cost sensors with $n_X = 10, 30, 50, 70$, respectively. In each case, three levels of batch size were selected, which correspond to 10\%, 50\%, and 100\% of the total sample size $n_X$. To assess the effect of batch size in adaptive spatial sampling, we obtained FRK predictions for each case of different sample size $n_X$ and batch size $b$. The model performance was summarized by calculating average prediction variance and the average root-mean-squared prediction error (RMSPE) over $D$, summarized in Figure \ref{fig:batchsize}. The results clearly indicate that both the average prediction error and RMSPE decrease as $n_X$ increases, but also the smaller batch size $b$ yields the lower value of average prediction error across all four different values of sample size $n_X$.

\end{document}